\documentclass[printer]{aa}  

\usepackage{graphicx}
\usepackage{txfonts}
\usepackage{xcolor}
\usepackage{natbib,twoopt}
\usepackage[breaklinks=true]{hyperref} 
\bibpunct{(}{)}{;}{a}{}{,} 
\makeatletter
\newcommandtwoopt{\citeads}[3][][]{\href{http://adsabs.harvard.edu/abs/#3}%
{\def\hyper@linkstart##1##2{}%
\let\hyper@linkend\@empty\citealp[#1][#2]{#3}}}
\newcommandtwoopt{\citepads}[3][][]{\href{http://adsabs.harvard.edu/abs/#3}%
{\def\hyper@linkstart##1##2{}%
\let\hyper@linkend\@empty\citep[#1][#2]{#3}}}
\newcommandtwoopt{\citetads}[3][][]{\href{http://adsabs.harvard.edu/abs/#3}%
{\def\hyper@linkstart##1##2{}%
\let\hyper@linkend\@empty\citet[#1][#2]{#3}}}
\newcommandtwoopt{\citeyearads}[3][][]%
{\href{http://adsabs.harvard.edu/abs/#3}
{\def\hyper@linkstart##1##2{}%
\let\hyper@linkend\@empty\citeyear[#1][#2]{#3}}}
\makeatother
\AtBeginShipout{%
  \ifnum\value{page}>1 %
    \typeout{* Additional boxing of page `\thepage'}%
    \setbox\AtBeginShipoutBox=\hbox{\copy\AtBeginShipoutBox}%
  \fi
}
\begin{document} 

   \title{VV~655 and NGC~4418: Implications of an interaction for the evolution of a LIRG \thanks{Based in part on observations made with the Southern African Large Telescope (SALT) as part of program 2014-2-SCI-052 (P.I.: J. S. Gallagher)}}

   \author{Erin Boettcher
          \inst{1,2}
          \and
          John S. Gallagher~III\inst{1}
          \and
          Youichi Ohyama\inst{3}
          \and
          Eskil Varenius\inst{4}
          \and
          Susanne Aalto\inst{5}
          \and
          Niklas Falstad\inst{5}
          \and
          Sabine K\"{o}nig\inst{5}
          \and
          Kazushi Sakamoto\inst{3}
          \and
          Tova M. Yoast-Hull\inst{6}
             }

   \institute{Department of Astronomy, University of Wisconsin-Madison, 475 N. Charter St., Madison, WI 53706, USA \\
                     \email{eboettcher@astro.uchicago.edu, jsg@astro.wisc.edu}
                     \and
                     {Department of Astronomy and Astrophysics, University of Chicago, 5640 S. Ellis Ave., Chicago, IL 60637, USA} 
                     \and
                     {Institute of Astronomy and Astrophysics, Academia Sinica, 11F of Astronomy-Mathematics Building, AS/NTU, No. 1, Sec. 4, Roosevelt Rd., Taipei 10617, Taiwan, R.O.C.}
                     \and
                     {Jodrell Bank Centre for Astrophysics, School of Physics and Astronomy, The University of Manchester, Alan Turing Building, Oxford Road, Manchester, M13 9PL, UK}
                     \and
                     {Department of Space, Earth and Environment, Onsala Space Observatory, Chalmers University of Technology, 439 92 Onsala, Sweden}
                     \and
                     {Canadian Institute for Theoretical Astrophysics, University of Toronto, 60 St. George St., Toronto, ON, M5S 3H8, Canada}
             }
   
   \date{Draft: \today}

 
  \abstract
      {VV~655, a dwarf irregular galaxy with HI tidal debris, is a companion to the lenticular luminous infrared galaxy (LIRG) NGC~4418. NGC~4418 stands out among nearby LIRGs due to its dense central concentration of molecular gas and the dusty, bi-polar structures along its minor axis suggestive of a wind driven by a central starburst and possible nuclear activity.}
      {We seek to understand the consequences of the ongoing minor interaction between VV~655 and NGC~4418 for the evolution of the LIRG. Specifically, we consider the origin of the gas supply responsible for the unusual nuclear properties of NGC~4418.}
      {We investigate the structural, kinematic, and chemical properties of VV~655 and NGC~4418 by analyzing archival imaging data and optical spectroscopic observations from the SDSS-III and new spectra from SALT-RSS. We characterize their gas-phase metal abundances and spatially resolved, ionized gas kinematics to better understand whether gas transfer between VV~655 and NGC~4418 resulted in the highly obscured nucleus of the LIRG.}
      {The gas-phase metallicity in NGC~4418 significantly exceeds that in VV~655. No kinematic disturbances in the ionized gas are observed along the minor axis of NGC~4418, but we see evidence for ionized gas outflows from VV~655 that may increase the cross-section for gas stripping in grazing collisions. A faint, asymmetric outer arm is detected in NGC~4418 of the type normally associated with galaxy-galaxy interactions.}
  {The simplest model suggests that the minor interaction between VV~655 and NGC~4418 produced the unusual nuclear properties of the LIRG via tidal torquing of the interstellar medium of NGC~4418 rather than through a significant gas transfer event. In addition to inducing a central concentration of gas in NGC~4418, this interaction also produced an enhanced star formation rate and an outer tidal arm in the LIRG. The VV~655-NGC~4418 system offers an example of the potential for minor collisions to alter the evolutionary pathways of giant galaxies.}


   \maketitle
%

\section{Introduction}
\label{sec_intro}

\object{VV 655} is a gas-rich dwarf irregular galaxy that is
interacting with \object{NGC 4418}, its lenticular luminous infrared
galaxy (LIRG) companion \citep{Armus87, Varenius17}. While
galaxy-galaxy interactions, such as major mergers, are widely studied
as important evolutionary factors, the diversity of interactions
between gas-rich dwarf and giant galaxies is yet to be systematically
explored. Even so, it is clear that minor mergers ($\lesssim$ 1:10
mass ratio) can fuel central activity in giant galaxies \citep[e.g.,
  see][]{Duprie96, Pisano99, Aalto01, Balcells01, Ferreiro04,
  Koribalski05, Sancisi08, Knierman10, Pearson18}. While simulations
and observations underscore the importance of minor mergers as drivers
of galaxy evolution, the impact of minor interactions that do not
involve mergers is less well studied, especially from an observational
perspective. These types of processes are difficult to detect,
especially in galaxies at cosmological distances, and can help to
explain the presence of LIRGs with apparently undisturbed structures
that are seen out to at least moderate redshifts
\citep[e.g.,][]{Pereirasantaella19}.

  NGC~4418 has several unusual nuclear properties, including a deep
  mid-infrared silicate absorption feature produced by cold dust
  \citep{Roche86, Spoon01, Roche15} and an extraordinarily dense
  (n(H$_{2}$) $= 10^{5} - 10^{7}$ cm$^{-3}$;
    \citealt{Costagliola15}) central concentration of
  $\sim$10$^{8}$~M$_{\odot}$ of molecular gas \citep{Kawara90,
    Aalto07, Lahuis07, Imanishi10, Sakamoto10, Gonzalez12,
    Costagliola15}, including an extremely Compton-thick nuclear
  region (N(H) $\gtrsim 10^{25}$ cm$^{-2}$;
    \citealt{Sakamoto13}). Kpc-scale bipolar dust filaments extend from
  the central regions of NGC~4418 \citep{Evans03, Sakamoto13,
    Ohyama19}, which contain a post-starburst stellar population as
  well as a possible highly obscured AGN \citep[][see also
    \S\ref{sec_pass}]{Roche86, Evans03, Shi05, Sakamoto13, Ohyama19}. These
  features suggest that an influx of interstellar matter reached the
  central zones of NGC~4418 within the past $\sim$Gyr, leading to
  enhanced star formation for at least the last 300~Myr \citep{Shi05,
    Ohyama19} and possibly fueling AGN activity.

The general properties of NGC~4418, including a centrally concentrated
power source, substantial molecular medium, and clear presence of
dust, are found in a significant fraction ($\sim$1/10 - 1/4) of
early-type field galaxies \citep[e.g.,][]{Koda05, Kaviraj12,
  Nyland17}. However, NGC~4418 stands out in terms of its status as a
LIRG, the extreme density of its circum-nuclear molecular medium, and
its ongoing interaction with VV~655, evidenced by the HI bridge
connecting the galaxies (\citealt{Varenius17}; see
Table~\ref{Tab:vv655_table} for a summary of the properties of these
galaxies). This presents the possibility that either a gas transfer
event or tidal torquing of the interstellar medium (ISM) of the LIRG
has shaped the nuclear properties of NGC~4418. A variety of studies,
including those focusing on early-type galaxies with molecular
interstellar matter by \citet{Lucero13} and \citet{OSullivan18},
emphasize the importance of minor mergers and interactions in
producing present-day examples of gas-rich, early-type
galaxies. NGC~4418 and VV~655 thus present a crucial case study for
better understanding the mechanisms by which minor interactions can
shape the evolutionary trajectories of gas-rich dwarf and giant galaxy
pairs.
    
    In this brief communication, we explore the connection between the
    minor interaction of NGC~4418 with VV~655 and the nuclear
    properties of the LIRG based on a combination of archival,
    multi-wavelength observations and a longslit spectrum obtained
    with the Southern African Large Telescope (SALT). The next section
    presents the observations, which we analyze in
    \S\ref{sec_ionized}. In particular, we address whether there is
    evidence for low-metallicity gas from VV~655 residing within
    NGC~4418. We also examine the ionized gas kinematics to search for
    evidence of gas accretion onto NGC~4418 and ionized gas outflows
    from VV~655 that could be feeding gas onto its companion. In
    \S\ref{sec_pass}, we explore possible connections between the
    interaction with VV~655 and the unusual properties of NGC~4418,
    including a comparison of the relevant
    timescales. \S\ref{sec_conc} contains a discussion of the results
    and our conclusions. We adopt a distance of 34~Mpc for the
    VV~655-NGC~4418 system, giving a scale of 160~pc~arcsec$^{-1}$.
  
\section{Observations}
\label{sec_obs}

\subsection{Spectroscopy}
\label{sec_spec}

We conducted our analysis using complementary spectroscopic data sets
from the Sloan Digital Sky Survey~III (SDSS-III; \S\ref{sec_SDSS}))
and SALT (\S\ref{sec_SALT}). As detailed below, the SDSS-III
single-fiber spectra provide broad wavelength coverage ($\lambda \sim
3800 - 9200$ \AA) and excellent flux calibration in the central
regions of the galaxies only. In contrast, the SALT longslit
observations offer spatially resolved spectroscopy across both
galaxies over a more limited wavelength range ($\lambda \sim 5560 -
6840$ \AA) at somewhat improved spectral resolution ($R \sim 2500$
vs. $2000$ near H$\alpha$).

\subsubsection{Data from the Sloan Digital Sky Survey~III}
\label{sec_SDSS}

%
   \begin{figure*}
   \centering \includegraphics[width=14cm]{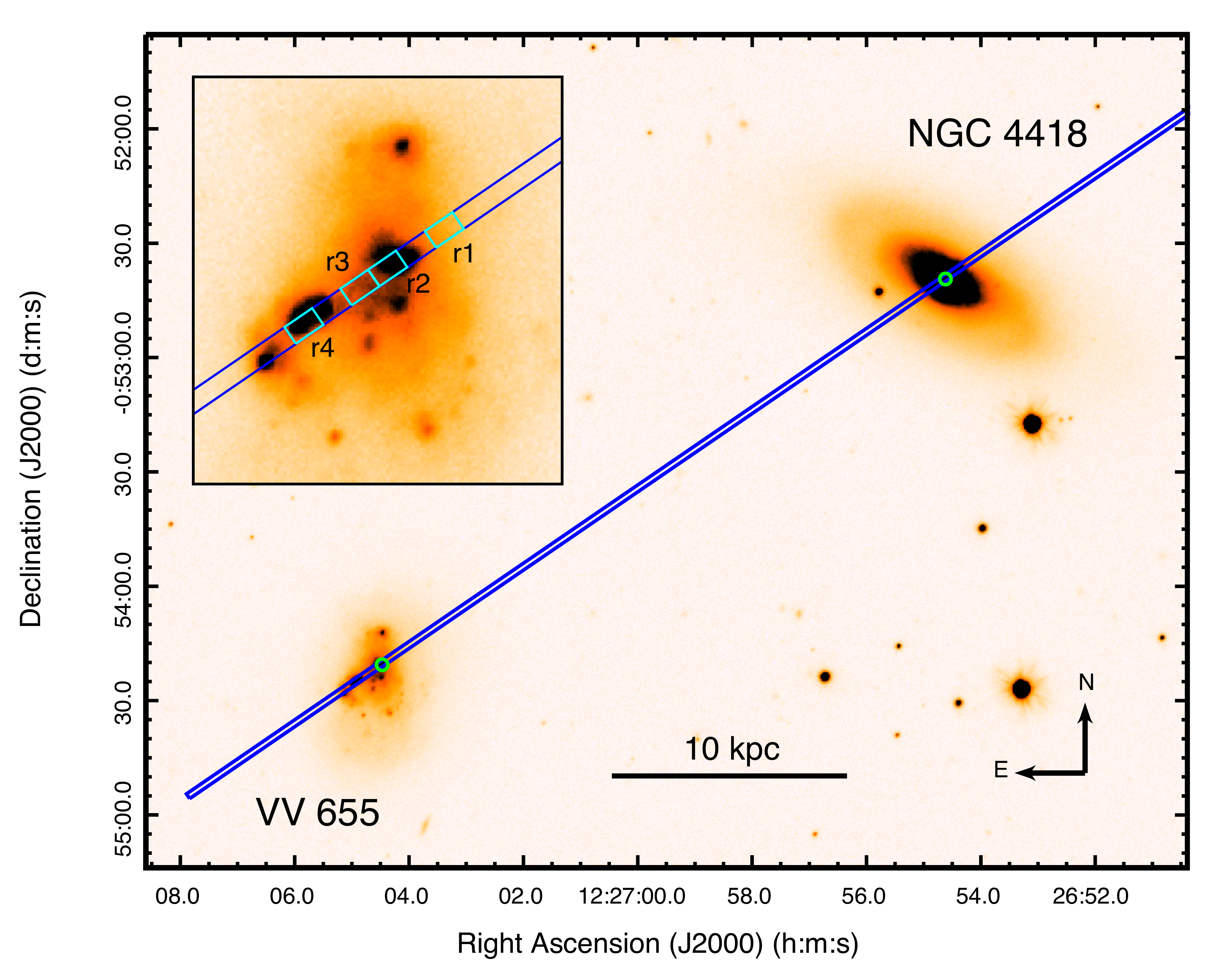}
      \caption{VV~655 and NGC~4418 illustrated in a reproduction of a
        KiDS DR4 \textit{g}-band archival image \citep{Kuijken19}; the
        galaxies are separated by a projected distance of
        $184''$ (30~kpc). The SALT-RSS longslit is shown in
        blue, and the SDSS fibers are indicated in green. The inset
        shows the locations of r1 - r4 on an enlarged image of
        VV~655. The SALT observations allow the gas-phase chemical
        abundances and any kinematic disturbances in the ionized gas
        to be characterized across the bodies of both galaxies.}
         \label{Fig:VV655_sys}
   \end{figure*}

The SDSS-III data release 12 (DR~12) Science Archive Server includes
reduced spectra of VV~655 (Object ID: 1237648720159965325, Plate:
0289, MJD: 51990, Fiber: 0202) and NGC~4418 (Object ID:
1237648720159965190, Plate: 0289, MJD: 51990, Fiber:
0208)\footnote{\url{http://skyserver.sdss.org/dr12/en/tools/explore/summary.aspx}}. These
spectra were taken with $3''$ diameter fibers whose locations are
shown in Fig.~\ref{Fig:VV655_sys}. In Table~\ref{Tab:sdss_table}, we
reproduce the emission-line ratios determined from the SDSS-III
Portsmouth spectroscopic re-analysis for both galaxies. The reduced
spectra extend redwards from approximately 3800~\AA \ and therefore
miss the [OII]$\lambda 3727$ doublet.

For VV~655, the ratio of I(H$\alpha$)/I(H$\beta$) $= 3.6$ observed by
SDSS is an upper limit since H$\beta$ emission is more strongly
affected by underlying stellar absorption than H$\alpha$ emission. As
a fiducial, we adopt an equivalent width for stellar absorption of
3~\AA\ for the stellar H$\alpha$ and H$\beta$ lines
\citep{Gallagher87}. Assuming that these add directly to the observed
line intensities gives I(H$\alpha$)/I(H$\beta$) $ = 3.0$, consistent
with the values from model fits in the SDSS \texttt{emissionLinesPort}
tabulation. This leads to an estimate of E(B-V) $\lessapprox$ 0.1~mag
for the internal reddening \citep[e.g.,][]{Garn10}, where the
foreground reddening is E(B-V) $ = 0.07$~mag. We do not perform a
reddening correction to the emission-line ratios discussed here due to
the small size of this correction and its uncertainty in a likely
complex, clumpy medium.

\subsubsection{Spectra from the Southern African Large Telescope}
\label{sec_SALT}

A longslit spectrum of VV~655 and NGC~4418 was taken on March 15, 2015
with the Robert Stobie Spectrograph (RSS) on SALT as part of program
2014-2-SCI-052 (P.I.: J. S. Gallagher). The $2\times 1250$~s
observations used a $1.5''$ slit width in combination with the
PG1800 VPH grating. We selected a slit position angle of 304.5 degrees
such that VV~655 was covered near the southeastern end of the
8-arcminute longslit as shown in Fig.~\ref{Fig:VV655_sys}. The slit
extends approximately along the minor axis of NGC~4418 and across
VV~655, thereby offering a direct, spatially resolved comparison of
their emission-line intensities and gas kinematics. In VV~655, the
SALT spectrum samples a 6~kpc swath across the galaxy, in contrast to
the 0.5~kpc sampled by the SDSS fiber that contains $\lesssim 10$\% of
the total light.

We used the PySALT science
pipeline\footnote{\url{http://pysalt.salt.ac.za/}} to perform a series
of corrections (bias, gain, and cross-talk) and to prepare and mosaic
the six frames that constitute each spectrum \citep{Crawford10}. We
then removed cosmic rays using the \texttt{L.A.Cosmic} task in
IRAF\footnote{IRAF is distributed by the National Optical Astronomy
  Observatories, which are operated by the Association of Universities
  for Research in Astronomy, Inc., under cooperative agreement with
  the National Science Foundation.}  \citep{vanDokkum01}. Using Ar
comparison lamp spectra, we determined the dispersion solution using
the \texttt{noao.twodspec.longslit.identify}, \texttt{.reidentify},
\texttt{.fitcoords}, and \texttt{.transform} tasks.

We used the IRAF task \texttt{images.immatch.imcombine} to perform a
median image stack with median scaling and weighting. To remove the
sky lines, we extracted and subtracted sky-line spectra from portions
of the slit free from object contamination. We calculated the
uncertainty per pixel by propagating the Poisson error through the
reduction process.

We extracted spectra along the regions of interest around VV~655 and
NGC~4418 using a median combination over an 11-pixel ($\sim 450$~pc)
aperture. We calculated the emission-line intensities by integrating
over the lines of interest in the continuum-subtracted spectra. The
NGC~4418 spectra show evidence of a broad stellar absorption feature
at H$\alpha$, which we fit and removed with an absorption line of
equivalent width $\sim 2.5$ \AA. The uncertainties on the measured
intensities account for the Poisson error and a 5\% error on the
continuum.

To characterize their kinematics, we fit the emission lines of
interest with single Gaussians using the IDL \texttt{gaussfit}
function. This function uses a least-squares approach to determine the
best-fit line centers, widths, and amplitudes and the uncertainties on
these parameters. In instances where two Gaussian components are
required to sufficiently represent the emission-line profile, we performed the
multi-component fit by minimizing the chi-squared statistic ($\chi^{2}
= \Sigma_{i = 0}^{N} \frac{(I_{\lambda,obs,i} -
  I_{\lambda,mod,i})^{2}}{\sigma_{\lambda,obs,i}^{2}}$) over a grid in
parameter space, where $I_{\lambda,obs,i}$ and $I_{\lambda,mod,i}$ are
the observed and modeled flux densities of the $i^{\text{th}}$ pixel
and $\sigma_{\lambda,obs,i}$ is the uncertainty on the former
quantity.

We thus characterized the emission-line intensities in instrumental
units, radial velocities, and velocity dispersions as functions of
position along the slit within both galaxies. All radial velocities
presented in this paper are heliocentric velocities determined using
the IRAF task \texttt{astutil.rvcorrect}. All line widths are
corrected for the instrumental resolution determined from the
comparison lamp spectra.

\subsection{Imaging and photometry}
\label{sec_image}

%
   \begin{figure}
   \centering
  \includegraphics[width=7.5cm]{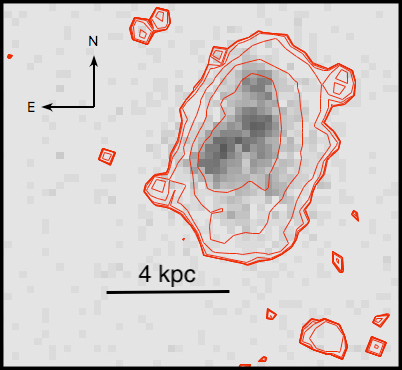}
      \caption{Structure of the stellar body of VV~655 shown
          by the GALEX far-UV image (grayscale) with overplotted
          contours from a Spitzer Space Telescope 3.6 $\mu$m band-1
          image. The outer isophote of VV~655 corresponds to
          $\approx$ 22.7~mag~arcsec$^{-2}$ ($\approx$
          10~L$_{\odot}$~pc$^{-2}$) and is relatively symmetric with
          no indication of tidal tails (see also
          Fig.~\ref{Fig:VV655_sys}).}
         \label{Fig:vv655_uv_36mu}
   \end{figure}

Archival images and photometry of VV~655 and NGC~4418 are available
from several sources; some of the derived properties of these galaxies
are summarized in Table~\ref{Tab:vv655_table}. As shown in
Fig.~\ref{Fig:vv655_uv_36mu}, the ultraviolet (UV) structure of VV~655
shows multiple UV-bright starforming regions, consistent with the
spatially extended emission lines observed in our SALT spectrum (see
\S\ref{sec_ionized}.) Thus, while the outer isophotes of VV~655 are
fairly regular, the star formation in this galaxy is found in
multiple, discrete clumps, suggestive of the irregular elliptical (iE)
class of blue compact dwarf galaxies \citep[e.g.,][]{Loose86}.

For VV~655, we measured integrated magnitudes from GALEX
far-UV (FUV) and Spitzer 3.6 $\mu$m band-1 images
(see Fig.~\ref{Fig:vv655_uv_36mu}). The Spitzer 3.6~$\mu$m luminosity
is $\sim$3$\times$10$^9$~L$_{\odot}$, which yields a stellar mass
estimate of a few $\times$10$^9$~M$_{\odot}$. For comparison, the gas
mass associated with VV~655 is $\sim$1$\times$10$^9$~M$_{\odot}$
\citep{Varenius17}, and thus despite the loss of a considerable gas
mass, VV~655 remains quite gas-rich.

The star formation rate of VV~655 derived from the FUV luminosity
using the calibration of \citet{Kennicutt12} is
$\sim$0.1~M$_{\odot}$~yr$^{-1}$, assuming E(B-V)~$= 0.07$. This result
is in good agreement with the star formation rate reported by
\citet{Varenius17}. Despite its prominent young stellar population and
substantial gas loss during the interaction with NGC~4418, VV~655
retains a sufficient gas supply to support star formation for
approximately an additional Hubble time and thus displays star
formation properties that are typical of dwarf irregular systems
\citep{Hunter85, vanZee01}.

   Since it is known that VV~655 and NGC~4418 are interacting
   \citep[e.g.,][]{Varenius17}, we investigated whether the stellar
   bodies of either galaxy show evidence for associated tidal
   distortions. The young stellar population of VV~655 as observed in
   the FUV is centrally concentrated, while the outer isophotes in the
   infrared are relatively symmetric with no clear indication of tidal
   tails. Deeper observations, however, are required to determine the
   degree to which the tidal debris from VV~655 lacks stars.

%
   \begin{figure}
   \centering
  \includegraphics[width=7.5cm]{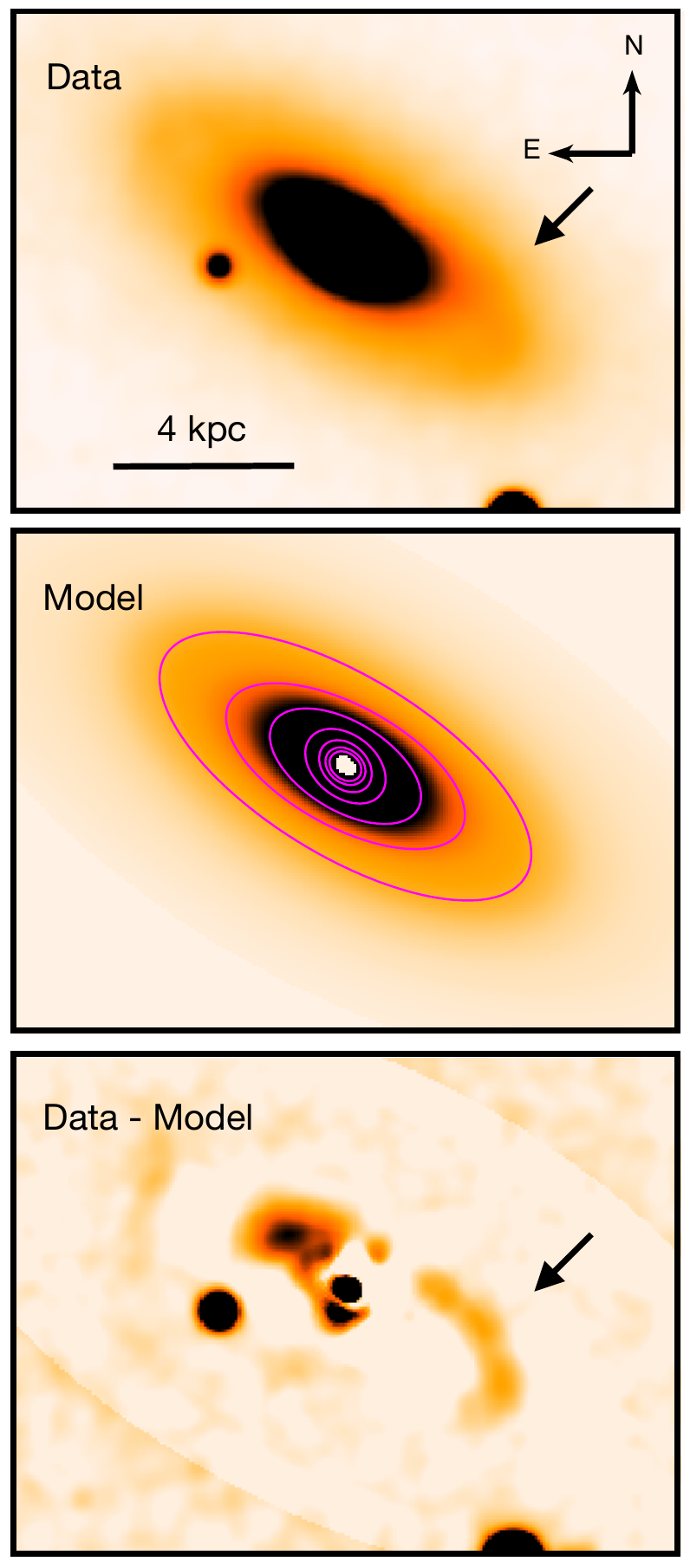}
      \caption{\textit{g}-band image of NGC~4418 from the KiDS archive
        \citep{Kuijken19} illustrating the internal structure of the
        outer disk before (top) and after (bottom) subtraction of a
        disk model (middle). The images are smoothed with a $\sigma =
        3$~pixels ($1.2''$) Gaussian kernel to improve the
        signal-to-noise ratio. The difference image reveals an
        asymmetric feature, indicated by the arrow, that has the
        properties of a one-sided tidal tail produced by a minor
        interaction. From the inside out, the contour levels in the
        middle panel are at 80\%, 60\%, 40\%, 20\%, 10\%, and 5\% of
        the surface brightness of the innermost contour. The central
        pixels in NGC~4418 are excluded from the model fit due to dust
        structures at this location. The features near the center of
        the galaxy in the bottom panel are due to undersubtraction of
        the smooth model from the centrally complex, clumpy data.}
         \label{Fig:n4418_tail}
   \end{figure}
   
Figure~\ref{Fig:n4418_tail} shows the \textit{g}-band KiDS image of
NGC~4418 \citep{Kuijken19} with a model of the galaxy disk subtracted
to display an arm-like feature on the southwest side of the galaxy. To
construct the disk profile, we made a smooth model of the light
distribution using the IRAF task \texttt{ellipse}. The bright star
that is projected against the galaxy to the east was effectively
removed by adopting the median option for deriving the isophote
levels. We used the bmodel capability in \texttt{ellipse} to produce a
symmetric model of the light distribution that we subtracted from the
original image to show the substructure. As the arm-like feature is
diffuse, we smoothed the original image with a $\sigma = 3$~pixels
($1.2''$) Gaussian kernel to improve the signal-to-noise ratio
before subtracting the model.

The arm-like structure revealed by this analysis is faint; its
contrast with respect to the surrounding disk is 5-10\% and it is
detected at the $5 - 10 \sigma$ level relative to the background after
model subtraction. The visible part of this feature, which we will
refer to as the ''outer arm'' (see below), contains $\sim$1\% of the
total \textit{R}-band flux from the galaxy. At a projected distance of
$\sim 4$~kpc from the center, the width of the outer arm is $\approx
6''$, or $\approx 1$~kpc. It is asymmetric in that a counterpart is
not detected on the opposite side of the galaxy. Since the stellar
rotation of the southwest side of NGC~4418 is redshifted \citep[see
  Fig. 6 of][]{Ohyama19} and extinction by a dusty polar outflow along
the northwest semi-minor axis suggests that the northwest side of the
galactic disk is the far side \citep{Sakamoto13, Ohyama19}, the
arm-like feature appears to be a trailing structure, as expected. For
this reason, we favor the interpretation of this feature as a disk
structure rather than a projected tidal stream in the halo.

The outer arm is also visible in the SDSS-III images, but its
presence to our knowledge has not been noted in the
literature. However, NGC~4418 has sometimes been classified as an Sa
galaxy with an outer ring. The presence of an outer ring is not
confirmed in modern data, but it is possible that the arm-like feature
in NGC~4418 was detected but incorrectly interpreted in some of the
earlier studies.\footnote{The \textit{Uppsala General Catalog of
    Galaxies} \citep{Nilson73} classifies NGC~4418 as an Sa galaxy and
  notes the presence of a superimposed companion that we suspect is
  the tidal arm feature.}

%

\begin{table*}
  \centering
  \caption{\normalsize{Properties of VV~655 and NGC~4418}}
  \label{Tab:vv655_table}
  \begin{tabular}{lllllc}
    \hline\hline                        
    \noalign{\smallskip}
     & & \textbf{VV~655} & & \textbf{NGC~4418} & \\
    \hline
    \noalign{\smallskip}
    Observed Property & Measured & Physical & Measured & Physical & Notes \\
    \hline
    \noalign{\smallskip}
        RA (J2000) & 12h27m04.56s & & 12h26m54.62s & & a \\
        Decl. (J2000) & -00d54m23.2s & & -00d52m39.4s & & a \\
        Redshift & 2202~km~s$^{-1}$ & 34~Mpc & 2106~km~s$^{-1}$ & 34~Mpc & b \\
        Far-UV Mag. & 16.8 & 4$\times$10$^{42}$erg~s$^{-1}$ & ... & ... & c \\
        3.6~$\mu$m Mag. & 12.5 & 2.4 $\times$10$^9$~L$_{\odot}$ & 9.9 & 2.7 $\times$10$^{10}$~L$_{\odot}$ & d \\
        HI Mass & 1$\times$10$^9$~M$_{\odot}$ & M(HI)/M$_*$  $\approx$ 1/2 & ... & ... & e \\
        $R_{P}(r)$ & $18''$ & 3.0~kpc & $27''$ & 4.3~kpc & f \\
        $v_{rot}$ & 20~km~s$^{-1}$ & 30-90 ~km~s$^{-1}$ & 130~km~s$^{-1}$ & 150~km~s$^{-1}$ & g \\
        SFR & & $\sim 0.1$ M$_{\odot}$ yr$^{-1}$ & & $\sim 3.2$ M$_{\odot}$ yr$^{-1}$ & h \\
        \hline
        \noalign{\smallskip}
  \end{tabular}
  \begin{flushleft}
       {\footnotesize {\it a}: From the NASA/IPAC Extragalactic
         Database (NED; \url{https://ned.ipac.caltech.edu/}). {\it b}:
         Systemic HI velocity from \citet{Varenius17}. {\it c}: From
         GALEX archival images, this paper; NGC~4418 is on the edge of
         the GALEX field of view and thus its FUV magnitude is not
         given. {\it d}: From Spitzer archival images, this
         paper. {\it e}: HI mass from \citet{Varenius17}, measured
         over the region shown in their Fig. 5; the HI content of
         NGC~4418 is not well determined due to HI absorption. Stellar
         mass from L$_{3.6}$ assuming M/L $= 1$. {\it f}: $r$-band
         Petrosian radius from SDSS-III DR 12. {\it g}: From
         \citet{Varenius17} and this paper (VV~655) and
         \citet{Ohyama19} (NGC~4418). {\it h}: From
         \citet{Varenius17}, as derived from the 1.4 GHz continuum,
         and this paper.}
  \end{flushleft}
\end{table*}
 
\section{Ionized gas}
\label{sec_ionized}

\subsection{Chemical abundances}

%

\begin{table*}
  \centering
  \caption{\normalsize{Emission-line ratios for VV~655 and NGC~4418}}
  \label{Tab:sdss_table}
  \begin{tabular}{lccccc}
    \hline\hline                        
    \noalign{\smallskip}
   Galaxy & [OIII]$\lambda 5007$/H$\alpha$ & [NII]$\lambda 6583$/H$\alpha$ & [SII]$\lambda 6716$/H$\alpha$ & [SII]$\lambda 6716$/[SII]$\lambda 6731$ & Notes \\
        \hline
        \noalign{\smallskip}
        \smallskip
         & & & \textbf{SDSS-III} & \\
        \smallskip
        VV~655 & $0.62 \pm 0.01$ & $0.098 \pm 0.004$  & $0.150 \pm 0.005$ & $1.43\pm 0.07$ & a \\
        NGC~4418 & $0.17 \pm 0.04$ & $0.95 \pm 0.05$ & $0.47 \pm 0.04$ & $1.2 \pm 0.1$ & a \\
        \smallskip
         & & & \textbf{SALT-RSS} & \\
        \smallskip
        VV~655 (r1; Diffuse) & ... & $0.121 \pm 0.007$ & $0.230 \pm 0.007$ & $1.52 \pm 0.08$ & \\
        VV~655 (r2: SF) & ... & $0.100 \pm 0.002$ & $0.170 \pm 0.003$ & $1.43 \pm 0.04$ & \\
        VV~655 (r3; Diffuse) & ... & $0.105 \pm 0.003$ & $0.163 \pm 0.003$ & $1.35 \pm 0.04$ & \\
        VV~655 (r4: SF) & ... & $0.091 \pm 0.001$ & $0.158 \pm 0.001$ & $1.43 \pm 0.02$ & \\
        NGC~4418 (Nucleus) & ... & $1.27 \pm 0.08$ & $0.42 \pm 0.06$ & $1.1 \pm 0.2$ & b \\
        NGC~4418 ($R = 7.5''$) & ... & $0.7 \pm 0.1$ & $0.45 \pm 0.09$ & $1.0 \pm 0.3$ & b \\
        \hline
  \end{tabular}
  \begin{flushleft}
    {\footnotesize {\it a}: Emission-line intensities and
      uncertainties are from the Portsmouth spectroscopic re-analysis
      presented on the SDSS-III DR 12 Science Archive Server
      (\url{http://skyserver.sdss.org/dr12/en/tools/explore/summary.aspx}.) 
      {\it b}: The continuum in the H$\alpha$ region of
      NGC~4418 is corrected for a broad stellar absorption line of
      strength $\sim 2.5$ \AA~(see Fig.~\ref{Fig:SALTspec}).}
  \end{flushleft}
\end{table*}

We quantify and compare the gas-phase metallicities in VV~655 and
NGC~4418 to explore the implications for gas transfer events(s)
between these galaxies. Our measured oxygen abundance for the ionized
gas in VV~655 is based on the strong emission-line abundance
calibration developed from SDSS spectra by \citet[][see their
  Fig.~9]{Curti17}. Using emission-line intensities for VV~655 from
the SDSS single-fiber spectrum, we find the index log(N$_2$) $= -1.01
\pm 0.02$, corresponding to [O/H] $= -0.3 \pm 0.1$ or about $40 -
60$\% of the solar value \citep{Asplund09}\footnote{N$_2$ is the
  intensity ratio between the [NII]$\lambda$6583 and H$\alpha$
  emission lines.}. The abundance based on the ratio of
I([OIII]$\lambda$5007) to I(H$\beta$) gives a similar result with an
abundance of $\sim$60\% of solar in the MPA-JHU galaxy properties
catalog.\footnote{\url{https://www.sdss.org/dr12/spectro/galaxy_mpajhu/}}

Using the SALT spectra, we quantify the spatially resolved
emission-line ratios in four regions of VV~655 - two star-forming (r2
and r4) and two diffuse regions (r1 and r3; see
Fig.~\ref{Fig:VV655_sys}). As reported in Table~\ref{Tab:sdss_table},
the [SII] doublet intensity ratio is close to the low-density limit,
indicating that most of the observed emission is coming from HII
regions and surrounding diffuse ionized gas. The low characteristic
value of log(N$_2$) $= -1.0 \pm 0.01$ and [NII] intensities below
those of [SII] agree with a metallicity approaching that of the Small
Magellanic Cloud (SMC) across the body of VV~655. Thus, both the
SDSS-III and SALT observations indicate that the gas-phase metallicity
in VV~655 is distinctly sub-solar, suggesting that this galaxy is not
a remnant of what was once a considerably more massive galaxy and is
unlikely to have been heavily disrupted by its interaction with
NGC~4418.

   \begin{figure*}
   \centering
  \includegraphics[width=16cm]{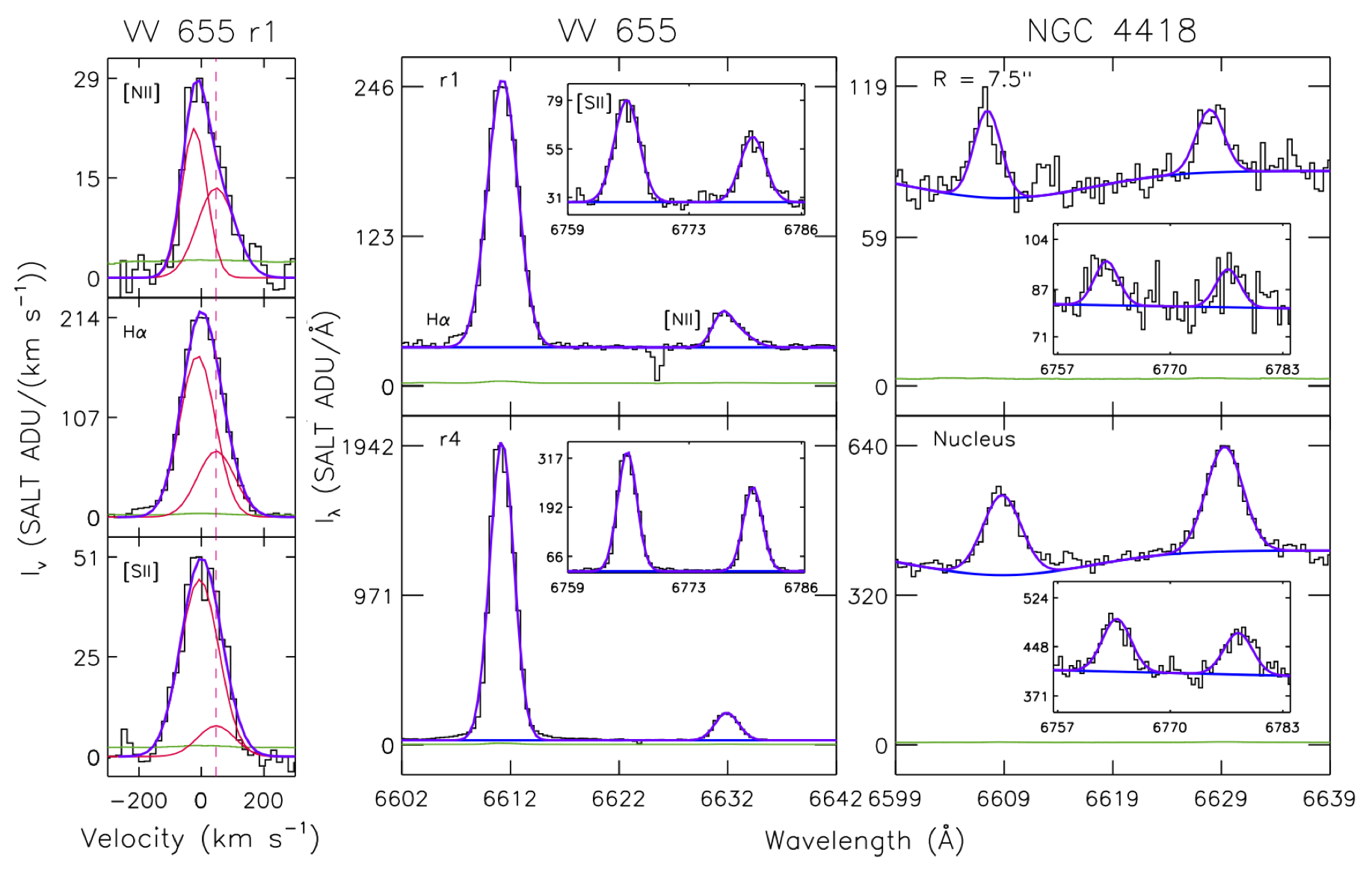}
      \caption{SALT-RSS spectra of emission lines in the H$\alpha$
        region from VV~655 (left) and NGC~4418 (right). The blue and
        purple lines indicate fits to the continuum and the single-
        and multi-component Gaussian emission-line profiles,
        respectively. The green curves indicate the error
        spectra. Spectra are shown for a diffuse (r1) and a
        star-forming region (r4) in VV~655 and near the nucleus and at
        an offset of $R = 7.5''$ (1.2 kpc) to the northwest in
        NGC~4418. The difference in the N$_{2}$ value of approximately
        an order of magnitude between the galaxies suggests that the
        metallicity of the outflowing, ionized gas in NGC~4418 exceeds
        that of the HII regions in VV~655. As shown at left, in the
        diffuse regions of VV~655, a redshifted [NII]$\lambda$6583
        wing extends beyond the velocity range where there are
        currently HI detections \citep{Varenius17}. We indicate upper
        limits on the corresponding kinematic components in the
        H$\alpha$ and [SII]$\lambda$6716 emission lines, which suggest
        that the redshifted component arises from gas with an N$_{2}$
        intensity ratio that exceeds the characteristic value for
        VV~655 by at least a factor of two. Thus, this feature likely
        indicates the presence of a shocked, ionized gas outflow. The
        wavelengths shown are the observed wavelengths. The feature
        near $\lambda = 6624$ \AA \ is an instrumental artifact.}
         \label{Fig:SALTspec}
   \end{figure*}

As shown in Fig.~\ref{Fig:SALTspec} and Table~\ref{Tab:sdss_table},
the emission-line ratios of the ionized gas in VV~655 markedly differ
from those of the extended ionized medium in its LIRG companion.  For
NGC~4418, we report SALT measurements at the location of the optical
continuum intensity peak (labeled nucleus) and at the largest distance
from the center at which a spectrum with sufficient signal-to-noise
ratio could be extracted ($R = 7.5''$ or 1.2 kpc to the northwest). It
is clear that the characteristic value of log(N$_2$) $= 0.0 \pm 0.04$
observed in NGC~4418 is a dex larger than that seen in VV~655; this
value is consistent with that found from the SDSS spectrum within the
uncertainties. The ionized gas along the south-southeast minor axis of
NGC~4418 in the direction towards VV~655 has emission-line ratios
consistent with ionized gas elsewhere in the galaxy.  \citet{Ohyama19}
also find N$_2 \gtrsim 1$ in the extended ionized medium of the
galactic outflow, and their independent analysis of the red emission
lines in NGC~4418 agrees with our conclusion that the chemical
abundance of the ionized gas is near solar values.
  
Since the material of greatest interest - that of the ionized outflow
from the nucleus of NGC~4418 - may be excited by shocks, chemical
abundances based on emission-line observations must be considered with
care. The optical spectra of supernova remnants in the SMC, where
shock excitation is an important factor, have high N$_2$ values above
those of HII regions but less than unity, and a similar result is also
found in the Large Magellanic Cloud \citep{Payne07,Payne08}. As the
shock-excited supernova remnants in the Magellanic Clouds bracket the
likely chemical abundances in VV~655 and have N$_2 < 1$, we conclude
that the ionized gas in NGC~4418 has higher metallicity than the HII
regions in VV~655.

\subsection{Ionized gas kinematics in NGC~4418}
\label{sec_ionized_4418}

  The velocity pattern traced by the red [NII] line in the NGC~4418
  SALT spectrum is consistent with that found by \citet{Ohyama19},
  with radial velocity slowly increasing along the minor axis from the
  northwest to the southeast. Although our SALT spectroscopy extends
  across the HI bridge connecting VV~655 and NGC~4418, we do not
  detect emission to appreciably larger distances toward VV~655 than
  Ohyama et al. ($\sim 10 - 15''$ or $1.6 - 2.4$ kpc toward the dwarf
  galaxy; see their Fig. 4). Neither work suggests strong evidence for
  a direct disturbance of the ionized gas envelope in NGC~4418
  associated with the HI tidal debris from VV~655. We do not detect
  emission lines at the location of the HI bridge between the two
  galaxies. As the tidal material is extended along an orbit that
  likely has at most moderate eccentricity, any gas donated from
  VV~655 must interact and lose angular momentum to reach the center
  of NGC~4418.

\subsection{Galactic outflow from VV~655?}
\label{sec_outflow}

Figure~\ref{Fig:SALTspec} shows a redshifted velocity wing in the
[NII]$\lambda$6583 emission line observed in the SALT spectra sampling
diffuse regions of VV~655. This feature is not seen in the H$\alpha$
or [SII] lines, indicating that it is a faint feature originating in
gas with a high N$_2$ ratio. It has a maximum offset of $\sim
100$~km~s$^{-1}$ from the peak emission, and it reaches to an average
observed velocity of $v_{\text{[NII]}} = 2305 \pm 10$~km~s$^{-1}$
across the body of VV~655. \citet{Varenius17} model the HI emission in
this galaxy with a blue and a red component; the latter is centered at
$v_{\text{HI}} = 2222.8$~km~s$^{-1}$ with a full width at half maxium
of $29.8$~km~s$^{-1}$, and thus HI is not detected at the maximum
[NII] velocity within the errors. While it is possible that HI with
column densities below the detection threshold is associated with the
redshifted wing, there is no evidence for an interaction between the
ionized wing and the HI in the present data.

In the left panel of Fig.~\ref{Fig:SALTspec}, we decompose the
[NII]$\lambda$6583 emission line observed in r1 into two Gaussian
components, one of which represents the redshifted wing. We find $N_2
\gtrsim 0.2$ for the high-velocity gas in both r1 and r3, and thus
this component arises from gas with an $N_2$ intensity ratio at least
a factor of two higher than is characteristic of VV~655, as often
occurs when gas is shock heated \citep[e.g.,][]{Payne07, Payne08,
  Newman14}. This combination of a redshifted velocity wing and
possible shock excitation is consistent with ionized gas outflows from
star-forming regions of dwarf galaxies. We note that these velocities may
also arise from chemically enriched, infalling gas associated with a
multi-phase galactic fountain flow or material stripped from VV~655
and/or NGC~4418 via their interaction.

Using the baryonic Tully-Fisher relationship and assuming an
inclination of $\gtrsim 45$~degrees based on the isophotal axis ratio,
we estimate a rotation speed $v_{rot} \approx 90$~km~s$^{-1}$ for
VV~655 given the calibration of \citet{Bell01}. Thus, if the
redshifted velocity wing is indicative of an outflow, the gas may
require assistance from tides and ram pressure to escape the galaxy,
potentially aiding the donation of mass and metals to NGC~4418
\citep[see][]{Bustard18}. Given the ongoing interaction and extended
tidal debris associated with VV~655, gas escape aided by the
interaction is a reasonable possibility \citep{McClureGriffiths2018},
although gas confinement by ram pressure may also occur
\citep{Marcolini03, Bustard18}.

\section{A close passage of VV~655 by NGC~4418: Consideration of timescales}
\label{sec_pass}

The HI maps of \citet{Varenius17} present the case for VV~655 to have
made a close passage by NGC~4418 $\sim 100 - 400$~Myr in the
past. This timescale is broadly consistent with the projected
separation of the two galaxies on the sky ($300$ kpc) and their
relative radial velocities ($\Delta$v $\approx 100$ km
s$^{-1}$). \citet{Varenius17} prefer a prograde model where the
orbital angular momentum vector for VV~655 is roughly aligned with the
rotation vector for NGC~4418. However, since the dynamical properties
of NGC~4418 were not known at the time, they do not exclude a
retrograde interaction. Transient, arm-like stellar density features
on one side of a disk galaxy can originate from such prograde, minor
interactions. For example, \citet{Helmi03} and \citet{Purcell11}
discuss how interactions with low-mass satellites can be responsible
for arm-like ring features in the outer disk of the Milky Way.

Following the discussion by \citet{Helmi03}, the limited angular
extent of the outer arm in NGC~4418 and its curvature suggest that
this is a relatively young feature. A tidal arm is expected to wrap up
and circularize over approximately a rotation time, or a few $\times
10^{8}$ years; it will then be disrupted by differential rotation over
several more rotation periods, suggesting an age of $<$1~Gyr. The
outer arm may have originated in several ways during an interaction,
including disturbance of disk material, as in the references above, or
as a remnant of material captured from VV~655
\citep[e.g.,][]{Mihos94}, though the regularity of the outer isophotes
of VV~655 at multiple wavlengths disfavor this scenario (see
Fig.~\ref{Fig:vv655_uv_36mu}). While the available data do not suffice
to fully illuminate the specific details of the dynamical event, the
age implied for the outer arm is consistent with its formation during
the close passage of VV~655.

\citet{Ohyama19} analyzed an SDSS optical spectrum of the central
regions of NGC~4418 to identify the components of its stellar
populations(s) \citep[see also][]{Shi05}. These authors show that the
stellar spectrum can be reproduced by combining a recent ``starburst"
stellar component with an age of $\sim$10~Myr with an older
``postburst" population with an age of $\sim$300-400~Myr that may
result from an earlier, interaction-induced starburst. Older
populations are also present that produce about 40\% of the
\textit{V}-band central zone light. Thus, the optically visible
stellar populations are suggestive of an event where the central star
formation rate increased several hundreds of Myr in the past and has
since remained active into the present, consistent with the close
passage of VV~655 both fueling the nucleus of the LIRG and forming the
outer arm. We note that optical data do not probe the deeply hidden
region around the nucleus where a significant fraction of the radiated
power is likely to originate, and thus do not offer reliable insights
into the properties of a centrally located active starburst in
NGC~4418.

Our analysis of nebular emission-line intensities establishes that the
ionized gas in VV~655 has substantially lower metallicity than that in
NGC~4418. The extended ionized medium in NGC~4418 cannot have come
recently from VV~655; any gas transferred to the LIRG must have been
in the galaxy for a sufficient time to be chemically enriched. A gas
residence time of a few 100~Myr in NGC~4418 would be sufficient to
enrich the gas in $\alpha$-elements and products of CNO nuclear
burning \citep[see, e.g.,][]{Matteucci01}. Thus, this time span
suffices for any ISM transferred from VV~655 at the time of closest
approach to be processed through multiple stellar generations and
enriched in chemical elements, consistent with the relative gas-phase
chemical abundances of VV~655 and NGC~4418 and the presence of dust in
the apparent outflow cone \citep[e.g.,][]{Evans03, Sakamoto13}.

\section{Discussion and conclusions}
\label{sec_conc}

The combination of a LIRG classification, post-starburst optical
spectrum, and unusual polar dust plumes indicate that NGC~4418 is in a
transitory evolutionary phase. These properties are generally
understood to result from an infusion of gas into the central regions
of galaxies. The resulting high-density ISM can support active star
formation and potentially fuel a central supermassive black hole. The
gas mass involved in the central activity in NGC~4418 is
$\geq$10$^{9}$~M$_{\odot}$ based on the masses of the young stars and
molecular gas \citep{Lisenfeld00, Ohyama19}, which sets a limit on the
nature of potential gas donors.

Interactions are the most likely triggers for gas infall events in
present day galaxies, and major mergers are well established to
generate dense central concentrations of molecular gas and associated
luminous central starbursts. NGC~4418, however, does not display the
prominent tidal tails, shells, or substantial asymmetries in its
stellar body that are associated with major mergers. A minor merger,
while capable of producing the observed behavior
\citep[e.g.,][]{Mihos94, Hernquist95}, is also unlikely. Such a merger
would have required the presence of a second, low-mass but gas-rich
companion galaxy in addition to VV~655 that has since merged with the
LIRG. While possible, this scenario has a low probability of
occurrance given the rarity of galaxies with two gas-rich satellites
that could supply the necessary $\sim$10$^{9}$~M$_{\odot}$ of ISM
\citep{Liu11, Tollerud11}.

A second possibility, considered by \cite{Mapelli15}, is for gas to be
transferred from VV~655 to NGC~4418. Models by \cite{Mapelli15}
indicate that gas capture from an orbiting satellite galaxy can occur
but will be inefficient. They estimate that about 10\% of the
satellite's gas can be transferred and potentially migrate to the
center of the primary system. In NGC~4418, this model suggests an
initial gas reservoir of $\approx 10^{10}$~M$_{\odot}$ in VV~655. This
exceeds the estimated HI mass associated with VV~655 from
\citet{Varenius17} by an order of magnitude. The lack of irregular
ionized gas kinematics along the minor axis of NGC~4418 and the need
for any accreted material to lose angular momentum to migrate to the
nuclear regions of the LIRG are also potential challenges to this
model.

The third scenario to consider is that the perigalacticon passage of
VV~655, likely on a prograde orbit, perturbed the pre-existing ISM in
NGC~4418 and led to its central concentration. A variety of models
show that a satellite with an SMC-like mass, such as VV~655, can
produce tidal torques that result in a significant inward migration of
the initial gas disk \citep{Pettitt16, RamonFox19}. Indeed, the
possibility of an interaction-induced inflow of molecular gas in the
nucleus of NGC~4418 was raised by \citet{Kawara90} and
\citet{Gonzalez12}. Studies of the HI content of early-type galaxies
reveal that gas masses in the range of 10$^{9}$~M$_{\odot}$ are
relatively common \citep[e.g.,][]{Noordermeer05, Morganti06,
  Oosterloo07}, with considerable circumgalactic reservoirs of cool
gas comparable to those found in late-type galaxies also often
associated with such systems \citep{Zahedy19}. It is thus entirely
plausible for NGC~4418 to have contained a substantial ISM prior to
its interaction with VV~655, consisting largely of HI supporting low
levels of star formation.

This class of models appears most likely to apply in the case of
NGC~4418 as it requires fewer special circumstances than either of the
other two scenarios and has the added advantage that the gas in an
early-type galaxy is expected to be at least moderately metal-rich and
dusty \citep{Griffith19}. This is in contrast to the possibility
discussed by \citet{Varenius17} in which the nucleus of NGC~4418 is
fed by ongoing accretion from VV~655's tidal debris. While it is
possible that a gas transfer event occurred during close passage and
the donated material has since been chemically enriched in NGC~4418,
the simplest model suggests that this is not the primary mechanism
responsible for the unusual nuclear properties of the LIRG.

  The innermost zones of NGC~4418, however, are deeply obscured in our
  optical observations. The ionized gas chemical abundances derived
  from optical data do not necessarily reflect conditions in the
  innermost nuclear region, unless gas is circulating between the
  nucleus and its surroundings. The presence of dusty plumes
  associated with a galactic outflow suggests that the ISM is unlikely
  to be quiescent and unmixed \citep[see][]{Ohyama19}; however, the
  degree to which mixing can occur is uncertain. Thus, it merits
  consideration whether the nucleus of NGC~4418 could be metal-poor
  due to recently accreted material from VV~655.

The rich molecular-line emission from the core of NGC~4418 is well
observed and allows us to test for dense, metal-poor
gas. Investigations of relative molecular abundances by
\citet{Lahuis07} and \citet{Costagliola15} emphasize the diversity of
the molecular content, including the prominence of nitrogen-bearing
molecules such as HCN. \citet{Sakamoto13} and \citet{Costagliola15}
also point to the similarity between emission in the nuclear region of
NGC~4418 and that from the massive merger Arp~220. These data indicate
that the dense nuclear region of NGC~4418 is not deficient in
$^{14}$N. This is a strong constraint since VV~655 is observed to have
a low abundance of this secondary element.

We conclude that a recent close passage between VV~655 and NGC~4418
affected the evolutionary trajectories of both galaxies. The unusual
nuclear properties of NGC~4418 most likely result from internal inward
migration of its ISM due to tidal torquing by VV~655 that occurred
$\sim$300~Myr in the past. This event led to the central starburst as
well as the likely ongoing substantial growth of the supermassive
black hole in NGC~4418 \citep[see][and references therein]{Sakamoto13,
  Roche15}. For its part, VV~655 is actively starforming; as
tabulated in the SDSS \texttt{emissionLinesPort} catalog, its
H$\alpha$ equivalent width of $\approx$ 80 \AA\ is consistent
  with a starburst, likely induced by the interaction. VV~655 has
clearly lost gas due to the interaction, yet retains low metallicity
in its ISM. The presence of a metal-enhanced outflow, possibly
stronger in the past, may partially account for the low metallicity in
this system.

In this scenario, the presence of extremely Compton-thick molecular
gas around the nucleus of NGC~4418 is occurring in the late phases of
its interaction with VV~655, $\sim$300~Myr after the peak of the
associated starburst event in NGC~4418. This process stands in
contrast to cases such as Arp~220, where extreme nuclear column
densities are associated with high-intensity starbursts in an ongoing
major merger. The NGC~4418 - VV~655 interaction therefore serves as a
signpost for the potential of minor interactions to produce profound
effects, including an extremely Compton-thick nucleus, long after the
peak of a collisionally induced disturbance from a low-mass
companion. This case study thus illustrates that minor interactions
can have considerable consequences for the more massive galaxy,
motivating further statistical studies and simulations to determine
the physical conditions under which the primary galaxy undergoes such
significant evolutionary effects.

Given the increase in galaxy interaction rates with redshift, it is
important to understand how various interaction scenarios shape the
ISM. The relative frequency of interactions between giant galaxies and
low-mass companions underscores the importance of characterizing such
systems at low redshift, where a spatially resolved understanding of
their stellar populations, morphologies, and gas kinematics and
chemical abundances inform the physics driving the evolution of
similar systems in the early universe.

\begin{acknowledgements}
Some of the observations reported in this paper were obtained with the
Southern African Large Telescope (SALT) as part of program
2014-2-SCI-052, P.I.: J. S. Gallagher. The continued support from
"Team SALT" is greatly appreciated. We thank the anonymous referees
for useful comments, and Hsiao-Wen Chen for helpful
discussions. Y. Ohyama and K. Sakamoto received support by the
Ministry of Science and Technology (MOST) of Taiwan, MOST
107-2119-M-001-026- (Y. O.) and MOST 107-2119-M-001-022- (K. S.). This
study would not have been possible without access to archival data
from the SDSS and KiDS, and we thank these teams for their
contributions to the astronomical research community. Funding for
SDSS-III has been provided by the Alfred P. Sloan Foundation, the
Participating Institutions, the National Science Foundation, and the
U.S. Department of Energy Office of Science. The SDSS-III web site is
http://www.sdss3.org/. SDSS-III is managed by the Astrophysical
Research Consortium for the Participating Institutions of the SDSS-III
Collaboration including the University of Arizona, the Brazilian
Participation Group, Brookhaven National Laboratory, Carnegie Mellon
University, University of Florida, the French Participation Group, the
German Participation Group, Harvard University, the Instituto de
Astrofisica de Canarias, the Michigan State/Notre Dame/JINA
Participation Group, Johns Hopkins University, Lawrence Berkeley
National Laboratory, Max Planck Institute for Astrophysics, Max Planck
Institute for Extraterrestrial Physics, New Mexico State University,
New York University, Ohio State University, Pennsylvania State
University, University of Portsmouth, Princeton University, the
Spanish Participation Group, University of Tokyo, University of Utah,
Vanderbilt University, University of Virginia, University of
Washington, and Yale University. Based on data products from
observations made with ESO Telescopes at the La Silla Paranal
Observatory under programme IDs 177.A-3016, 177.A-3017, 177.A-3018 and
179.A-2004, and on data products produced by Target/OmegaCEN,
INAF-OACN, INAF-OAPD and the KiDS production team, on behalf of the
KiDS consortium. OmegaCEN and the KiDS production team acknowledge
support from: Deutsche Forschungsgemeinschaft, ERC, NOVA and NWO-M
grants, Target, the University of Padova, and the University Federico
II (Naples). Members of INAF-OAPD and INAF-OACN also acknowledge
support from the Department of Physics \& Astronomy of the University
of Padova, and of the Department of Physics of Univ. Federico II
(Naples). This publication makes use of data products from the
Wide-field Infrared Survey Explorer, which is a joint project of the
University of California, Los Angeles, and the Jet Propulsion
Laboratory/California Institute of Technology, funded by the National
Aeronautics and Space Administration. This work is based in part on
observations made with the Spitzer Space Telescope and the Galaxy
Evolution Explorer, which are operated by the Jet Propulsion
Laboratory, California Institute of Technology, under contract with
the National Aeronautics and Space Administration. This work has made
use of NASA's Astrophysics Data System and the NASA/IPAC Extragalactic
Database (NED) and Infrared Science Archive which are operated by the
Jet Propulsion Laboratory, California Institute of Technology, under
contract with the National Aeronautics and Space Administration.
\end{acknowledgements}

\bibliographystyle{aa} 
\bibliography{vv655_final.bib}  

\end{document}